\documentclass[useAMS,usenatbib]{mn2e}

\usepackage{verbatim} 
\usepackage{natbib} 
\usepackage{amsmath} 
\usepackage{amsbsy}
\usepackage{amssymb}
\usepackage{mathrsfs} 
\usepackage{lscape} 
\usepackage{graphicx}
\usepackage{epstopdf}
\usepackage{deluxetable}

\newcommand{\Lya}{${\rm Ly}\alpha$}

\title[Small Cold Streams Covering Factor]{The Small Covering Factor of Cold Accretion Streams}
\author[Claude-Andr\'e Faucher-Gigu\`ere and Du\v{s}an Kere\v{s}]{Claude-Andr\'e Faucher-Gigu\`ere\thanks{Miller Fellow; cgiguere@berkeley.edu} and Du\v{s}an Kere\v{s}\thanks{Hubble Fellow}\\
Department of Astronomy and Theoretical Astrophysics Center, University of California, Berkeley, CA 94720-3411, USA.}

\begin{document}
\maketitle


\begin{abstract}
Theoretical models of galaxy formation predict that galaxies acquire most of their baryons via cold mode accretion. 
Observations of high-redshift galaxies, while showing ubiquitous outflows, have so far not revealed convincing traces of the predicted cold streams, which has been interpreted as a challenge for the current models. 
Using high-resolution, zoom-in smooth particle hydrodynamics simulations of Lyman break galaxy (LBG) halos
combined with ionizing radiative transfer, we quantify the covering factor of the cold streams at $z=2-4$. 
We focus specifically on Lyman limit systems (LLSs) and damped Ly$\alpha$ absorbers (DLAs), which can be probed by absorption spectroscopy using a background galaxy or quasar sightline, and which are closely related to low-ionization metal absorbers. 
We show that the covering factor of these systems is relatively small and decreases with time. 
At $z=2$, the covering factor of DLAs within the virial radius of the simulated galaxies is $\sim3$\% ($\sim1\%$ within twice this projected distance), and arises principally from the galaxy itself. 
The corresponding values for LLSs are $\sim10$\% and $\sim4$\%. 
Because of their small covering factor compared to the order unity covering fraction expected for galactic winds, the cold streams are naturally dominated by outflows in stacked spectra. 
We conclude that the existing observations are consistent with the predictions of cold mode accretion, and outline promising kinematic and chemical diagnostics to separate out the signatures of galactic accretion and feedback. 
\end{abstract}


\section{INTRODUCTION}
\label{intro}
Recent years have seen the emergence of a new theoretical paradigm for galaxy formation, in which galaxies are predicted to acquire most of their baryons via cold, filamentary streams penetrating deep inside dark matter halos (e.g., Katz et al. 2003; Kere\v{s} et al. 2005, 2009a)\nocite{2003ASSL..281..185K, 2005MNRAS.363....2K, 2009MNRAS.395..160K}. 
This prediction from cosmological hydrodynamical simulations, in addition to being supported by idealized 1D simulations and analytic arguments \citep[][]{2003MNRAS.345..349B, 2006MNRAS.368....2D}, has now been confirmed by several groups using both smooth particle hydrodynamics (SPH) and grid-based codes \citep[e.g.,][]{2008MNRAS.390.1326O, 2009ApJ...694..396B, 2009Natur.457..451D}. 
Understanding how galaxies get their gas is a fundamental problem for many reasons. 
Observationally, measurements of the star formation rate density across cosmic time and of the neutral atomic and molecular reservoirs show that galaxies must continuously accrete material from the ionized intergalactic medium (IGM) in order to sustain their star formation \citep[][]{2009ApJ...696.1543P, 2010ApJ...717..323B}. 
Theoretically, the mode of gas accretion determines the rate at which baryons can settle at the bottom of dark matter halos, and therefore how rapidly galaxies can grow as a function of redshift and host halo properties \citep[e.g.,][]{baryona}. 
Whether baryons flow in galaxies in the form of narrow, dense streams or via the cooling of quasi-spherical atmospheres is also directly relevant to the efficiency of feedback processes, and therefore potentially to the quenching of star formation in massive galaxies and the establishment of the red sequence (e.g., Dekel \& Birnboim 2006; Kere\v{s} et al. 2005, 2009b\nocite{2006MNRAS.368....2D, 2005MNRAS.363....2K, 2009MNRAS.396.2332K}).
Finally, the cold streams may be connected to several observed phenomena, including high-redshift clumpy discs 
\citep[e.g.,][]{2006ApJ...650..644E, 2006Natur.442..786G}, Ly$\alpha$ blobs \citep[but see][]{2010ApJ...725..633F}, and high-velocity clouds around local galaxies \citep[e.g.,][]{2009ApJ...700L...1K}.
\begin{figure*}
\includegraphics[width=1.0\textwidth]{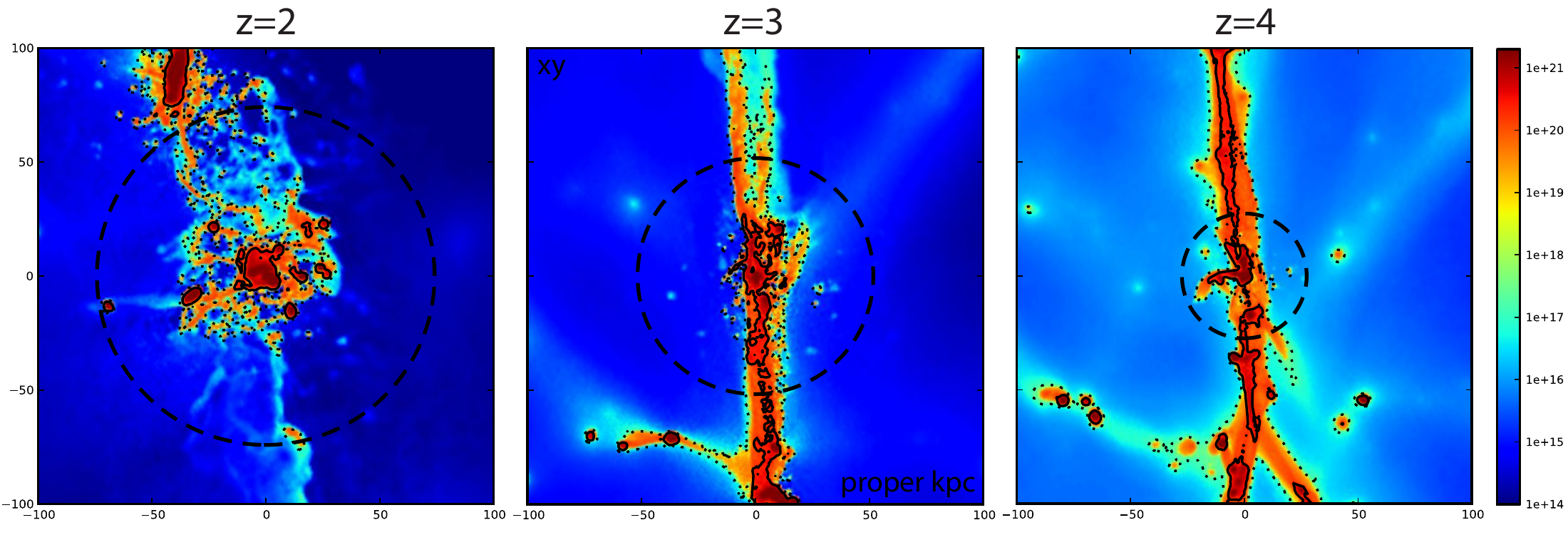}
\caption[]{Projected neutral hydrogen column density (xy projection in Table \ref{covering factor table}; cm$^{-2}$) around the B1 Milky Way progenitor at $z=2,~3,$ and 4. 
The dotted contours indicate Lyman limit systems, while the regions corresponding to damped \Lya~absorbers are shown by the solid contours. 
The point deepest in the potential is at the center of each panel and the dashed circles mark the virial radii. A gas smoothing length of 27 proper pc is achieved at $z=2$, making this one of the highest-resolution simulations of such a halo to date.}
\label{B1 maps}
\end{figure*}

In spite of the key role the cold mode is likely to play in galaxy formation and evolution, observations have so far provided little direct evidence of infalling cold halo gas. 
Thanks to novel techniques, including the use of projected quasar-quasar \citep[e.g.,][]{2009ApJ...690.1558P}, quasar-galaxy \citep[e.g.,][]{2005ApJ...629..636A, 2006ApJ...637..648S} and galaxy-galaxy pairs \citep[e.g.,][]{2010ApJ...717..289S}, observers have recently made impressive progress in probing the halo gas content of high-redshift galaxies. 
While the spectroscopic absorption measurements show strong evidence for ubiquitous galactic outflows at $z\sim2-3$, they do not show clear signatures of the predicted accretion streams, which has been interpreted as a challenge for current models \citep[e.g.,][]{2010ApJ...717..289S}. 
Exactly what these observations tell us about the cold mode has however been ambiguous, as we are only beginning to quantify how the predictions for the accretion dynamics should manifest themselves in observables. 
The opportunity that is presented to us by this state of affairs can hardly be understated, as it is a unique case of a nearly \emph{ab initio} physical prediction in galaxy formation that has yet to be tested, but for which observations are rapidly approaching truly discriminating power. 
It therefore behooves us to robustly quantify the predictions of our models. 

In \cite{2010ApJ...725..633F}, we made a step in this direction by calculating the Ly$\alpha$ cooling emission from cold streams, showing that the treatment of self-shielded gas introduced orders-of-magnitude uncertainties in most previous work, and concluding that pure accretion cooling is unlikely to dominate the Ly$\alpha$ emission of the bright Ly$\alpha$ blobs now routinely imaged \citep[e.g.,][]{2010arXiv1010.2877M}, although it could be important for some fainter sources \citep[e.g.,][]{2008ApJ...681..856R}. 
In this \emph{Letter} we address absorption spectroscopy as a different observational probe of the cold streams. 
Using high-resolution cosmological simulations of a Milky Way progenitor and of a typical LBG-mass halo, 
we quantify the covering factor of the cold streams in Lyman limit (LLSs) and damped \Lya~systems (DLAs). 
We show that the covering fraction of such dense gas decreases with time and that it is sufficiently small at $z\sim2$ that the presence of cold streams in the halos of Lyman break-selected galaxies is consistent with current observations, consisting principally of stacked spectra, that are dominated by outflow signatures.  

\begin{table*}
\label{covering factor table}
\centering
\caption{Covering Factors of Cold Mode Lyman Limit Systems and Damped Ly$\alpha$ Absorbers Around the B1 Milky Progenitor}
\begin{tabular}{ccccccccc}
\hline\hline
z  & $M_{\rm h}$ & $R_{\rm vir}$ & LLS($<0.5R_{\rm vir}$) & DLA($<0.5R_{\rm vir}$) & LLS($<R_{\rm vir}$) & DLA($<R_{\rm vir}$) & LLS($<2R_{\rm vir}$) & DLA($<2R_{\rm vir}$)  \\
\hline
  & $M_{\odot}$ & prop. kpc & \% (xy/xz/yz) & \% (xy/xz/yz) & \% (xy/xz/yz) & \% (xy/xz/yz) & \% (xy/xz/yz) & \% (xy/xz/yz) \\

\hline
2 & $3.2\times10^{11}$ & 74 & 32 / 40 / 23 & 10 / 9 / 10 & 10 / 15 / 11 & 3 / 4 / 3 & 4 / 4 / 4 & 1 / 1 / 2 \\
3 & $2.5\times10^{11}$ & 52 & 33 / 23 / 45 & 17 / 34 / 17 & 16 / 16 / 23 & 6 / 12 / 5 & 8 / 5 / 10 & 3 / 3 / 2 \\
4 & $7.4\times10^{10}$ & 27 & 50 / 29 / 51 & 25 / 53 / 25 & 32 / 27 / 32 & 11 / 25 / 10 & 17 / 15 / 14 & 6 / 13 / 6 \\
\hline
\end{tabular}
\tablenotetext{}{The three covering factors given for each case correspond to projections along different Cartesian axes (on the xy, xz, and yz planes).}
\end{table*}

\section{SIMULATIONS AND RADIATIVE TRANSFER}

\subsection{Hydrodynamics}
\label{hydro}
Our simulations use a modified version of the GADGET 2 cosmological code \citep[][]{2005MNRAS.364.1105S}. 
The gas dynamics is calculated using an SPH algorithm that conserves both energy and entropy \citep[][]{2002MNRAS.333..649S}.
The modifications with respect to the public version of the code include the treatment of cooling, the effects an uniform ultra-violet background (UVB), and a multiphase star formation algorithm as in \citet[][]{2003MNRAS.339..289S}.
The thermal and ionization properties of the gas are calculated including all relevant processes in a plasma with primordial abundances of hydrogen and helium following \cite{1996ApJS..105...19K}, using an update of the \cite{1996ApJ...461...20H} UV background including galaxies and quasars.  
Self-shielding effects are modeled in post-processing, as described in the next section. 
To achieve ultra-high resolution, we `zoom in' on individual halos within a $(10/h$ comoving Mpc$)^{3}$ simulation box and  follow the local gas dynamics at a resolution refined $64\times$ in mass. 

Our main simulation of a Milky Way progenitor (labeled B1) is identical to the one analyzed by \cite{2009ApJ...700L...1K}, but with a mass resolution higher by a factor of 3.375 because we used $192^{3}$ dark matter particles in the unrefined volume instead of $128^{3}$. 
The minimum achieved gas smoothing length is 27 proper pc at $z=2$, with a Plummer-equivalent gravitational force softening of $92 [(1+z)/3]^{-1}$ proper pc. 
The halo mass $\approx3\times10^{11}$ M$_{\odot}$ at $z=2$ corresponds to a low-mass LBG, and the simulation is among the highest-resolution cosmological SPH realizations of such a halo to date. 
The dark matter particle mass of this simulation is $2\times10^{5}$ M$_{\odot}$, and gas particle mass is $4\times10^{4}$ M$_{\odot}$. 
We also study a zoom-in simulation of a more typical LBG halo (labeled A1) of mass $\approx9\times10^{11}$ M$_{\odot}$ at this redshift, with a spatial resolution lower by a factor of 1.5. 
We assume a flat $\Lambda$CDM cosmology with $\Omega_{\rm m}=0.27$, $\Omega_{\rm b}=0.044$, $h=0.7$, $\sigma_{8}=0.8$, and $n_{\rm s}=0.95$, consistent with the latest \emph{WMAP} analysis \citep[][]{2010arXiv1001.4538K}. 
Galactic winds are neglected in this work to isolate the accretion streams, but will be studied in future work. 
We define the virial radius, $R_{\rm vir}$, as the radius enclosing a mean overdensity of 180 times the mean matter density.

\subsection{Radiative Transfer}
\label{RT}
The covering factor of the cold streams can only be meaningfully quantified for specific boundary criteria. 
For absorption statistics a natural definition is tied to the column density of the material. 
We limit our attention here to hydrogen, and in particular to the dense LLSs ($10^{17.2} \leq N_{\rm HI} \leq 2\times10^{20}$ cm$^{-2}$) and DLAs ($N_{\rm HI} > 2\times10^{20}$ cm$^{-2}$) that are often associated with proto-galaxies and their outskirts. 
By definition, these systems can shield themselves from the ionizing radiation of the cosmic background and local sources. 
To model the ionization balance, we use the ray tracing code introduced in \cite{2010ApJ...725..633F}. 
The UV background is modeled by following $512^{2}$ rays originating from the 6 faces of a cubic volume of side length 1/h comoving Mpc centered on the halo of interest. 
We have improved this code for the present work by including frequency-dependent attenuation of the radiation field, and by allowing for the possibility of enhanced ionizing flux from local sources by radially casting rays outward. 

Assuming that the central galaxy emits ionizing photons in proportion to its star formation rate, for a Salpeter IMF and for an escape fraction of 5\%, we find that ionization from local sources has a very small impact on the covering factor of surrounding LLSs, and is completely negligible for DLAs. 
This is easy to understand, as the optical depth of DLAs to ionizing photons exceeds $10^{3}$, so that the bulk of their gas does not see changes in ionizing luminosity unless they are extreme. 
Because the covering factors of LLSs and DLAs are insensitive to local ionization, we chose not to include it in the results presented here, which has the benefit of ensuring that we are providing upper bounds (in the absence of additional physics), and therefore strengthens our main point that they are small for the cold streams, especially at $z\sim2$. 
To avoid artificial collisional ionization of the interstellar material (ISM) due to the effective temperatures carried by the multiphase particles, we assume that these are fully neutral, up to a maximum $n_{\rm HI}=1$ cm$^{-3}$ to model the conversion to H$_{2}$ \citep[e.g.,][]{2001ApJ...562L..95S}.

\section{Results}
\label{results}
Figure \ref{B1 maps} shows the projected HI column density distribution around the simulated B1 Milky Way progenitor at $z=2,~3,$ and 4. 
Table \ref{covering factor table} summarizes the properties of the system at these redshifts and quantifies the covering factors of LLSs and DLAs. 
At each redshift, we provide the covering factors averaged within 0.5, 1, and 2 virial radii, and for three mutually orthogonal projections labeled according to the Cartesian projection plane (xy, xz, or yz), where the xy projection is the one shown in Figure \ref{B1 maps}. 
The covering factors are evaluated exactly by considering all the pixels of the projected radiative transfer grids. 
Providing the covering factors normalized to fractions of the virial radius gives a simple way to scale the results to halos of different masses. 
However, the availability of background sources in observations is independent of the virial radius of the foreground galaxy. 
In a forthcoming study, we will also provide numbers as a function of absolute distance for samples representative of observations, but it is already clear from Figure \ref{B1 maps} that much of the covering factor evolution within fixed fractions of the virial radius is driven by the evolution of the virial radius itself, and that the covering factors within a fixed distance do not evolve as strongly. 
For reference, the covering factors of DLAs [LLSs] between fixed radii of 10 and 100 proper kpc, for the xy projection of the B1 halo, are $(1,~2,~3)$\% [$(7,~8,~10)$\%] at $z=(2,~3,~4)$. 

At $z=4$, where the streams are prominent and the virial radius (27 proper kpc) is compact, $10-25$\% of the area within $R_{\rm vir}$ is covered by a DLA, depending on the sightline. 
By $z=2$, however, this fraction is reduced to $3-4$\% and even so most of it arises from galactic material rather than from the accretion streams, which are seen to actively fragment (excising the inner 10 proper kpc in radius, the DLA covering factor drops to $1-3$\%, depending on the sightline). 
Within 2 $R_{\rm vir}$, the DLA covering factor plummets below 1\% even including the galaxy. 
The filament fragmentation in this high-resolution simulation of a $\approx3\times10^{11}$ M$_{\odot}$ halo at $z=2$ is consistent with previous, lower-resolution work that showed that streams survive mostly in halos below the transition mass $\approx2-3\times10^{11}$ M$_{\odot}$, or at $z>2$. 
At lower redshift, cooling and possibly Rayleigh-Taylor instabilities start operating in the infalling gas, which is increasingly surrounded by a hot medium, and seed the formation of cool clouds as in \cite{2009ApJ...700L...1K}.

\begin{figure}
\begin{center}
\includegraphics[width=0.425\textwidth]{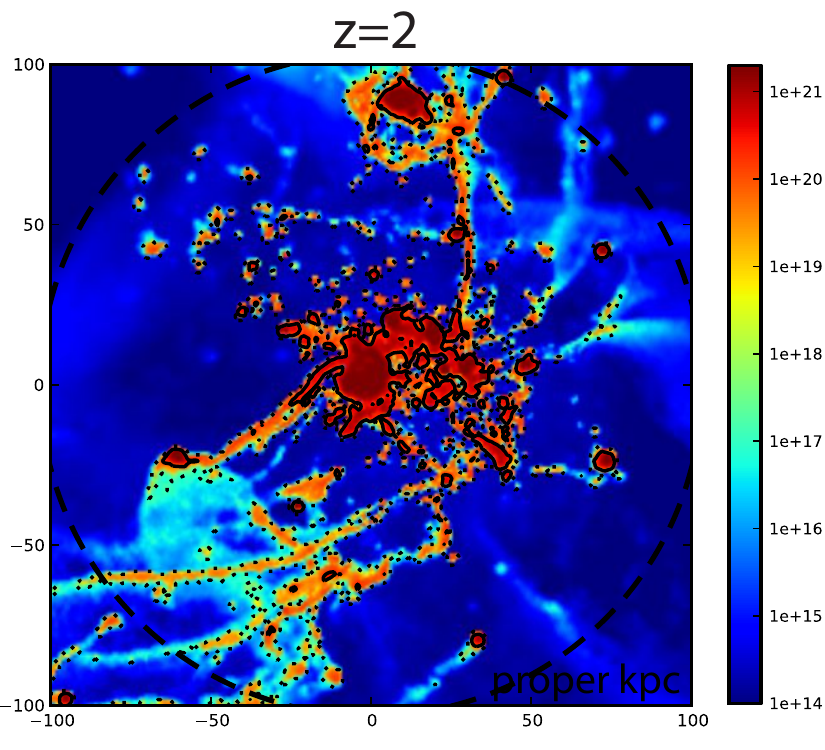}
\end{center}
\caption[]{Same as Figure \ref{B1 maps}, but for the A1 halo of mass $9\times10^{11}$ M$_{\odot}$ at $z=2$, corresponding to an average-mass LBG in the \cite{2010ApJ...717..289S} sample.}
\label{A1 map z2}
\end{figure}

Figure \ref{A1 map z2} shows the HI column density map around the A1 halo, of mass $\approx9\times10^{11}$ M$_{\odot}$ at $z=2$. 
While the resolution of this simulation is poorer by 50\% spatially, the halo mass is exactly the average mass probed by the LBG sample of \cite{2010ApJ...717..289S} at the mean redshift $\langle z \rangle=2.2$, and allows us to test how the results above scale to this mass. 
For the random projection shown, the covering factor of DLAs within $R_{\rm vir}$ at $z=2$ is 4\%, i.e. the same as for the xz projection of the B1 halo in Figure \ref{B1 maps}, and again arises mostly from material close to the galaxy. 
The LLS covering factor of 11\% within $R_{\rm vir}$ is also similar to the B1 case. 
This result is consistent with the study of \cite{2010arXiv1008.4242H}, who find that the DLA covering factor within $R_{\rm vir}$ in simulations without outflows is relatively constant over a fairly wide mass interval in the LBG range. 
The small covering factors of accreting material we find are therefore generic predictions for LBGs at this redshift. 
In simulations with outflows, \cite{2010arXiv1008.4242H} show that the DLA covering factor increases in high-mass halos, but their kinematics suggest that most of the increase owes to absorption by wind ejecta. 
Our point regarding the accretion streams is therefore unaffected to first order, although it is conceivable that interactions with outflows might affect their dynamics somewhat. 
Between the fixed radii of 10 and 100 proper kpc, the covering factors of DLAs and LLSs are almost identical, at $4$\% and $12$\% respectively, but higher than the corresponding numbers in the B1 case owing to the overall difference in halo size.

\section{DISCUSSION}
\subsection{Comparison with Observations}
A key question is whether the existing observations are consistent with cold accretion. 
While we have quantified the covering factors of LLSs and DLAs around $z=2-4$ galaxies, direct observational measurements of these quantities are not yet available in most cases. 
Our comparison is therefore somewhat qualitative at this time, as more accurate inferences will require larger simulation samples and the explicit modeling of metal absorbers. 
LLSs and DLAs are however closely connected to low-ionization metal species such as OI, SiII, CII, FeII, and MgII, which have ionization potentials similar to HI \citep[e.g.,][]{1990ApJS...74...37S, 2005ARA&A..43..861W}. 
It is also clear from from Figures \ref{B1 maps} and \ref{A1 map z2} that the column densities of neutral gas drop sharply outside the self-shielded regions. 

The most exhaustive constraints on the circum-galactic medium of $z\sim2-3$ galaxies were recently reported by \cite{2010ApJ...717..289S}, based on spectroscopic observations of a large number of LBGs, including 512 close angular galaxy-galaxy pairs. 
While both the blueshifted interstellar absorption and the redshifted \Lya~emission in direct sightlines provide strong evidence for high-velocity outflows, these authors concluded that there is at present little evidence of infalling cool material. In fact, they showed that a simple outflow model calibrated from direct sightlines also reproduces the observed transverse statistics in galaxy pairs. 
Unlike background quasars, the vast majority of LBGs are too faint to yield spectra of sufficient resolution and signal-to-noise ratio to study individual absorbers. 
The main results of \cite{2010ApJ...717..289S} are consequently based on stacked measurements. 
The authors also argue that the principal metal lines they consider are strongly saturated, so that the observed equivalent widths are good proxies for their covering factors and insensitive to metallicity. 
If that is the case, then infalling material would leave negligible traces in composite spectra of transverse absorption whenever its covering factor is much smaller than that of outflows. 
Since the mean redshift of Steidel et al.'s foreground galaxies is $\langle z \rangle=2.2$, this hypothesis is well supported by the small cold stream covering factors we predict at $z\sim2$. 
As shown by \cite{2010arXiv1008.4242H}, outflows can in contrast fill a substantial fraction of the virial radius with self-shielding systems.   
Note that we do not confirm the prediction of a larger, $\sim25$\% covering factor of DLAs between radii 20 and 100 proper kpc predicted by \cite{2009Natur.457..451D} for $10^{12}$ M$_{\odot}$ halos at $z=2.5$. The simulation analyzed by those authors had a spatial resolution than lower than ours by a factor $\sim15$ at this redshift, and they did not perform a radiative transfer calculation but instead included all the gas with temperature $T<10^{5}$ K in their column densities. Both effects go in the direction of overestimating the covering factors.

The above argument does not apply to direct single sightlines, in which velocity offsets $v_{\rm los}\geq0$ might be more directly associated with infall and in which absorption is most likely to occur close to the center. 
However, \cite{2010arXiv1012.0059K} recently demonstrated that the infalling material (with velocities of order the halo  circular velocity, $v_{\rm circ}$) is in general completely masked by absorption by the rotating ISM of the galaxy itself. 
With regards to Ly$\alpha$ emission, \cite{2010ApJ...725..633F} showed that accretion via cold streams (again because of their small covering factor, which allows the photons to take paths of least resistance between them) can only produce a minor blue enhancement. 
Winds with larger covering factors therefore drive the radiative transport and yield redshifted lines. 

We stress that the cold accretion paradigm is consistent with the presence of galactic winds. 
We in fact fully expect the real high-redshift galaxies to drive powerful winds, as appears required to explain a number of observations beyond those discussed above, including the low-end of the galaxy luminosity function and the pollution of the IGM with metals. 
The omission of galactic winds in the simulations analyzed here simply reflects the fact that, whereas simulating accretion from the cosmic web is a well-defined problem, we do not yet know how to robustly model winds. 
We have therefore deliberately excluded them to isolate the signatures of the cold streams. 
This is a reasonable approximation, as the small covering factor of the dense cold streams implies that it is difficult for the more dilute winds to strongly affect them \citep[e.g.,][]{baryona}.

\subsection{Inflows vs. Outflows Diagnostics}
The present lack of direct observational evidence for cold accretion and the small predicted stream covering factors, especially relative to the ubiquitous galactic winds at high redshift, highlight the challenge of testing the theory. 
We conclude by outlining ways in which future studies might be able to reveal the cold accreting gas. 

The most direct diagnostics of accretion versus winds are the kinematics and chemical composition of cool absorbers. 
In transverse spectra, the direction of motion of the absorbers is usually ambiguous because their physical positions are unknown.
However, the $\emph{magnitude}$ of the velocity offset provides a discriminant, as material infalling from the IGM should have a velocity $v_{\rm circ}\approx245{\rm~km~s^{-1}}
\left( \frac{M_{h}}{10^{12}~{\rm M_{\odot}}} \right)^{1/3}
\left( \frac{1+z}{3} \right)^{1/2}$, whereas outflows are observed to have velocities up to 800 km s$^{-1}$ or more \citep[e.g.,][]{2010ApJ...717..289S}. 
A caveat is that wind ejecta can in principle fall back onto galaxies with velocities $\sim v_{\rm circ}$, if their launch velocity is modest or if the galactic potential well is sufficiently deep \citep[e.g.,][]{2010MNRAS.406.2325O}. 
Metallicity may then be the only means to convincingly associate a particular absorber with intergalactic accretion. 
By carefully modeling the absorbers in their high signal-to-noise, high-resolution quasar spectra, \cite{2006ApJ...637..648S} were able to derive the metallicity of individual systems near galaxies, and link them to feedback processes based on their near solar abundances. 
By contrast, the IGM metallicity is $\sim1/700$ of solar at $z=2.5$ \citep[e..g,][]{2004ApJ...606...92S}, although material infalling onto galaxies could be substantially more pre-enriched owing to dwarf galaxies in filaments, and it is an important question for future modeling exactly what are the typical and maximum metallicities expected for cold mode accretion.

A practical issue is that most of the information currently available on absorbers within the halos of LBGs is from close galaxy-galaxy pairs \citep[][]{2010ApJ...717..289S}. 
Because galaxies are faint, it is in general not possible to model individual absorbers and infer their metallicity. 
The holy grail would be a large sample of close quasar-galaxy pairs from which to build up the kinematic and metallicity statistics of halo absorbers, but the rarity of quasars limits the usefulness of this approach. 
Deep spectroscopy of gravitationally lensed galaxies would also be very helpful, and next generation 30-m class telescopes will greatly improve the resolution and signal-to-noise ratio of galaxy spectra, and therefore enhance the information accessible via the galaxy-galaxy technique. 
Sufficiently sensitive integral field spectrographs would be ideal instruments to map the extended gas distribution around LBGs.

These observational proposals are however challenging, and underscore the critical role to be played by more detailed theoretical modeling. 
In analogy to how statistics from a large number of galaxy-galaxy pairs partially make up for the rarity of suitable quasar-galaxy pairs, more statistical predictions will connect our models better with the existing data. 
For example, it should be possible to theoretically compute the average equivalent width of different ions versus impact parameter, as measured by \cite{2010ApJ...717..289S} from galaxy-galaxy pairs, and use this information to constrain the models. 
For this exercise to be fruitful, it will however be necessary to improve our modeling to consistently include the galactic winds that produce most of the measured absorption, the attendant chemical enrichment, and to simulate a sizable sample of halos at sufficient resolution. 
We plan to address these problems in future work, as well as to investigate the constraints that can be derived on high-redshift outflows \citep[][]{abscosmo}. 
In the mean time, it is becoming clear that the signatures of cold accretion are quite subtle, and that accurate predictions will be needed to make sense of the data. 

\section*{Acknowledgments}
We thank Volker Springel for providing the enhanced version of his GADGET code, and acknowledge useful discussions with Lars Hernquist, Chung-Pei Ma, Matt McQuinn, Eliot Quataert, Alice Shapley, Rob Simcoe, and Chuck Steidel.  
CAFG is supported by a fellowship from the Miller Institute for Basic Research in Science. 
DK is supported by NASA through Hubble Fellowship grant HST-HF-51276.01-A. 
The computations presented in this paper were performed on the Odyssey cluster at Harvard University.

\bibliography{references} 
 
\end{document}